\documentclass[lettersize,journal]{IEEEtran}
\usepackage{amsmath,amsfonts}
\usepackage{algorithmic}
\usepackage{algorithm}
\usepackage{array}
\usepackage[caption=false,font=normalsize,labelfont=sf,textfont=sf]{subfig}
\usepackage{textcomp}
\usepackage{stfloats}
\usepackage{url}
\usepackage{verbatim}
\usepackage{graphicx}
\usepackage{cite}
\usepackage{balance}
\usepackage{multirow}   
\usepackage{xcolor}
\usepackage[T1]{fontenc}
\usepackage{newtxtext}
\begin{document}

\title{SAFT: Sensitivity-Aware Filtering and Transmission for Adaptive 3D Point Cloud Communication over Wireless Channels}

\author{Huda~Adam~Sirag~Mekki,
        Hui~Yuan,~\IEEEmembership{Senior Member,~IEEE},
        Mohanad~M.~G.~Hassan,
        Zejia~Chen,%
        \text{ and } Guanghui~Zhang
\thanks{This work supported in part by the National Natural Science Foundation of China under Grants 62222110, 62571303 and 62172259, the Taishan Scholar Project under Grant tsqn202103001, and the Shandong Provincial Natural Science Foundation under Grant ZR2022ZD38. (Corresponding author: Hui Yuan)}
\thanks{Huda~Adam~Sirag~Mekki, Hui Yuan, Mohanad~M.~G.~Hassan, and Zejia~Chen are with the School of Control Science and Engineering, Shandong University, Ji’nan, 250061, China, and also with the Key Laboratory of Machine Intelligence and System Control, Ministry of Education, Ji’nan, 250061, China (e-mail: hudasirag28ce@gmail.com; huiyuan@sdu.edu.cn; mohanad@mail.sdu.edu.cn; sduczj@mail.sdu.edu.cn).}
\thanks{G.~Zhang is with the School of Computer Science and Technology, Shandong University, Qingdao, China (e-mail: gh.zhang@sdu.edu.cn).}}

\markboth{Journal of \LaTeX\ Class Files,~Vol.~14, No.~8, August~2021}%
{Shell \MakeLowercase{\textit{et al.}}: A Sample Article Using IEEEtran.cls for IEEE Journals}


\maketitle

\begin{abstract}
Reliable transmission of 3D point clouds over wireless channels is challenging due to time-varying signal-to-noise ratio (SNR) and limited bandwidth. This paper introduces sensitivity-aware filtering and transmission (SAFT), a learned transmission framework that integrates a Point-BERT-inspired encoder, a sensitivity-guided token filtering (STF) unit, a quantization block, and an SNR-aware decoder for adaptive reconstruction. Specifically, the STF module assigns token-wise importance scores based on the reconstruction sensitivity of each token under channel perturbation. We further employ a training-only symbol-usage penalty to stabilize the discrete representation, without affecting the transmitted payload. Experiments on ShapeNet, ModelNet40, and 8iVFB show that SAFT improves geometric fidelity (D1/D2 PSNR) compared with a separate source--channel coding pipeline (G-PCC combined with LDPC and QAM) and existing learned baselines, with the largest gains observed in low-SNR regimes, highlighting improved robustness under limited bandwidth.
\end{abstract}

\begin{IEEEkeywords}
Point cloud transmission, robustness-aware transmission, deep joint source-channel coding, SNR adaptation
\end{IEEEkeywords}

\section{Introduction}\label{sec:intro}
\IEEEPARstart{P}{oint clouds} have become a mainstream 3D data representation in cutting-edge technologies, including autonomous driving, digital twin infrastructure, remote sensing, and immersive communication \cite{ref1_pointnet}. Accordingly, learning-based point cloud coding has also been studied for efficient delivery under practical system constraints \cite{Guarda2021MM, Ascenso2023MM}. In particular, their ability to represent complex geometric details as unordered sets of 3D coordinates makes point clouds well-suited to reconstructing physical environments \cite{ref2_pami_survey3dpc}. However, point clouds often contain many points to capture fine geometry, leading to large data volumes and substantial storage and transmission overhead. This makes reliable point-cloud delivery particularly challenging over wireless links with limited bandwidth or varying noise levels \cite{ref3_tip_graph_pcc, ref4_tcstv_teleimmersion}.

Legacy point-cloud transmission systems typically adopt a two-stage pipeline consisting of a source codec (e.g., geometry-based point cloud compression (G-PCC) or video-based point cloud compression (V-PCC)) and subsequent channel coding (e.g., LDPC, turbo, or polar codes) \cite{ref5_iso_23090_9_gpcc}. Although this separation-based approach works under favorable conditions, it can degrade sharply under severe noise or tight bandwidth constraints, where transmission errors may cause decoding instability and noticeable quality loss \cite{ref6_jetcas_mpeg_pcc}. To address these limitations, recent research has explored deep joint source-channel coding (DeepJSCC), where neural networks learn compression and channel robustness jointly by mapping input point clouds directly to transmission symbols\cite{ref7_tccn_deepjscc_images}.

Despite their promise, many DeepJSCC methods for point clouds still treat point tokens as equally important for reconstruction and adopt a fixed latent structure without explicit adaptation to communication constraints \cite{ref8_icassp_pcst,ref9_tvt_semcom_pointcloud}. In addition, robustness can be limited when the signal-to-noise ratio (SNR) varies at test time, since models are often trained for a narrow range of channel conditions. 
To address these limitations, we propose sensitivity-aware filtering and transmission (SAFT), an adaptive point-cloud transmission model for noisy wireless channels.
Our encoder adopts a Point-BERT-inspired point-token transformer that represents local point patches as tokens and models their relations via self-attention~\cite{ref22_cvpr_pointbert}.
The proposed method includes a token-level sensitivity module that learns token-wise scores to reduce reconstruction distortion under channel perturbation.
During training, a symbol-usage penalty stabilizes the discrete representation without affecting the transmitted payload.
Finally, the decoder is conditioned on a receiver-side SNR estimate for SNR-aware reconstruction. The main contributions of this work are the following:
\begin{enumerate}
    \item We propose a robustness-aware point-cloud transmission model that integrates a Point-BERT-inspired encoder with an SNR-aware decoder. This design extracts contextual geometric representations and conditions reconstruction on the decoder-side SNR  estimation.
    \item We introduce sensitivity-based token filtering module for robustness-aware transmission under channel perturbation.
    \item We design a multi-term training objective (sparsity, diversity, and training-only symbol-usage penalty) to stabilize the discrete latent space without transmitting side information.
\end{enumerate}

The remainder of this paper is organized as follows. In Section~\ref{sec:related}, we describe the related work briefly. In Section~\ref{sec:system}, we introduce the proposed method in detail. Experimental results and analyses are given in Section~\ref{sec:experiments}. Finally, we conclude the paper in Section~\ref{sec:conclusion}.

\begin{figure*}[t]
  \centering
  \includegraphics[width=0.95\textwidth,trim=0 0 0 0,clip]{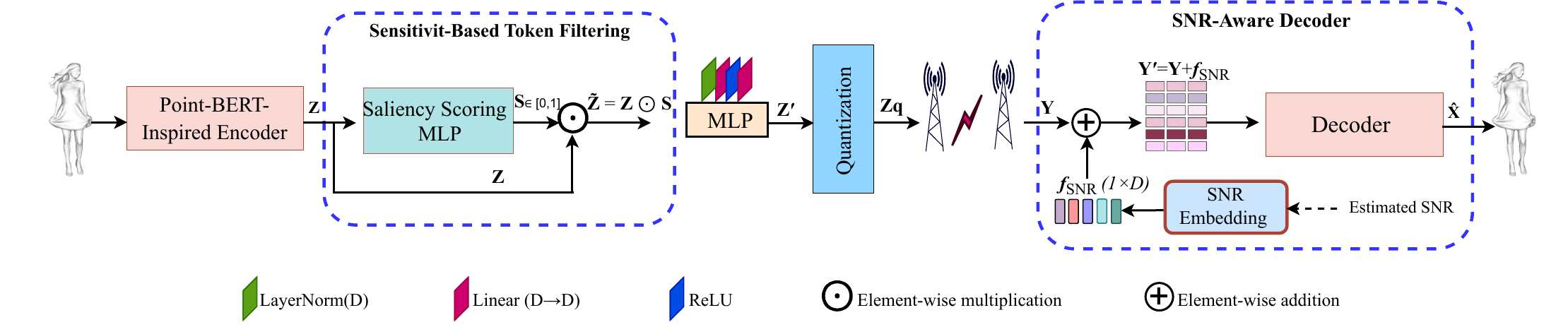}
  \caption{Overview of the proposed SAFT.}
  \label{fig:system_model}
\end{figure*}
\section{Related Work}\label{sec:related}
 In recent years, DeepJSCC has been widely studied for robust transmission over noisy wireless channels. Early efforts focused on 2D image transmission, while more recent work has explored extending learned communication to structured 3D data such as point clouds.
This section reviews representative developments in DeepJSCC for images and overviews recent work on DeepJSCC for 3D point-cloud transmission.

\subsection{DeepJSCC for Images}\label{sec:images}
Prior work has introduced several mechanisms to improve robustness and adaptability in learned JSCC for images. 
DeepSC-RI \cite{ref10_arxiv_deepsc_ri} employs a dual-branch design and cross-attention fusion to mitigate channel distortions and semantic inconsistency, while D2-JSCC \cite{ref12_pimrc_d2jscc} integrates semantic compression with digital channel coding to reduce end-to-end distortion. 
Building on these ideas, PADC \cite{ref13_twc_padc} introduces variable-length DeepJSCC with a quality predictor and dynamic control to adapt to changing conditions. 
Beyond architecture-level changes, related studies investigate adaptive quantization, region-dependent processing, and SNR-aware enhancement \cite{ref11_dcn_gan_semantic,ref15_jsac_image_semantic_coding,ref17_arxiv_snr_semantic_image,ref19_tccn_swinjscc}. 
Vector-quantized and goal-oriented latent representations have also been explored for task-driven communication \cite{ref20_wcl_vq_semcom,ref21_arxiv_goal_oriented_semantic}.
Although these methods demonstrate effective adaptation mechanisms for 2D signals, extending them to point clouds remains challenging due to irregular geometry, variable sampling density, and token-level sensitivity to channel perturbations.
\subsection{DeepJSCC for 3D Point Clouds}\label{sec:pointclouds}
Compared with images, relatively few studies investigated DeepJSCC for 3D point clouds. The lack of a regular grid structure and the variability of sampling density make efficient representation and robust transmission challenging. SEPT \cite{ref35_spawc_wireless_pc} reduces bandwidth usage by combining feature extraction with downsampling and incorporates SNR adaptation, but aggressive downsampling can sacrifice fine geometric details.
More recently, \cite{ying_icc2025_jsccm_pointcloud} proposed joint semantic--channel coding and modulation for point clouds by mapping features to digital constellations and allocating variable-length symbol sequences with channel-adaptive modulation, i.e., modulation-level adaptation.
SemCom \cite{ref30_globecom_semcom_pc} employs dual encoding for global and local features, while PCST \cite{ref8_icassp_pcst} introduces progressive resampling and sparse convolution for DeepJSCC under AWGN and Rayleigh fading channels. TSCS \cite{ref36_iwrfat_tscs_pointcloud} adopts a transformer encoder--decoder to mitigate cliff effects, but does not provide explicit control of the transmitted token count.
In parallel, learned point cloud compression has advanced rapidly toward high-fidelity visual perception; For instance,~\cite{zhou_tip2025_structure_pcc} presents a structure-aware generative framework that progressively reconstructs point-cloud surfaces to improve perceptual quality and coding efficiency.
While recent methods have explored channel-adaptive modulation and rate–distortion objectives, explicit token-level payload ratio control under channel perturbations remains less explored. 

\section{System Model}\label{sec:system}
The proposed sensitivity-aware filtering and transmission (SAFT) aims to ensure robust delivery of 3D point clouds over noisy wireless channels. As illustrated in Fig.~\ref{fig:system_model}, SAFT first encodes the input point cloud into patch-level latent tokens using a Point-BERT-inspired transformer. Next, a sensitivity-guided token filtering (STF) unit assigns a learned score to each token and filters token representations via continuous weighting to reduce reconstruction distortion under channel perturbation. The filtered tokens are then quantized and power-normalized for wireless transmission. Finally, at the receiver, an SNR-aware decoder reconstructs the point cloud from the received tokens conditioned on a receiver-side SNR estimate.
\begin{figure*}[!t]
  \centering
  \makebox[\textwidth][c]{%
    \includegraphics[width=.75\textwidth,clip,trim=0 40 160 0]{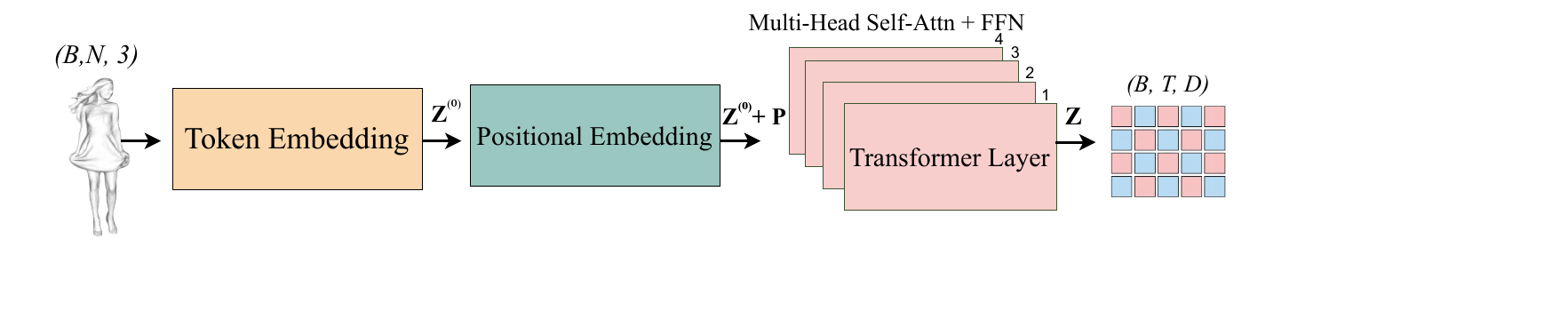}}
  \caption{Detailed architecture of the transformer-based point-token encoder.}
  \label{fig:encoder_arch}
\end{figure*}

\subsection{Encoder Architecture}\label{sec:encoder}
Given an input point cloud $\mathbf{X}\in\mathbb{R}^{B\times N\times 3}$, the encoder maps it to a sequence of $T$ patch-level tokens for transformer processing, as shown in Fig.~\ref{fig:encoder_arch}. Following Point-BERT-style tokenization \cite{ref22_cvpr_pointbert,ref23_iccv_pointtransformer}, each token is embedded by a learnable linear projection:
\begin{equation}
\mathbf{Z}^{(0)}=\text{Linear}_{3\rightarrow D}(\mathbf{X}), 
\end{equation}
where $D$ is the feature dimension (we use $D=256$). To provide positional cues for attention, we add learnable positional embeddings $\mathbf{P}\in\mathbb{R}^{1\times T\times D}$ shared across the samples:
\begin{equation}
\mathbf{Z}^{(0)} \leftarrow \mathbf{Z}^{(0)}+\mathbf{P}.
\end{equation}
Here, $\mathbf{P}$ provides sequence-order cues for attention and does not replace the geometry-dependent token features extracted from $\mathbf{X}$.

The token sequence is then processed by $L$ transformer encoder layers, each composed of multi-head self-attention (MHSA) and a feedforward network (FFN) with residual connections:
\begin{equation}
\begin{aligned}
\mathbf{Z}^{(\ell+1)}
&=\mathbf{Z}^{(\ell)}+\text{FFN}\Big(\text{LayerNorm}\big(\mathbf{Z}^{(\ell)} \\
&\qquad\qquad +\,\text{MHSA}(\text{LayerNorm}(\mathbf{Z}^{(\ell)}))\big)\Big),
\end{aligned}
\end{equation}
for $\ell=0,\ldots,L-1$. The FFN is implemented as a two-layer MLP:
\begin{equation}
\text{FFN}(\mathbf{u})=\text{Linear}_{512\rightarrow D}\!\Big(\text{ReLU}\big(\text{Linear}_{D\rightarrow 512}(\mathbf{u})\big)\Big).
\end{equation}
After the encoder stack, the latent tokens $\mathbf{Z}\in\mathbb{R}^{B\times T\times D}$ are forwarded to the STF and quantization blocks for transmission.

\begin{figure}[!t]
  \centering
  \includegraphics[width=\linewidth]{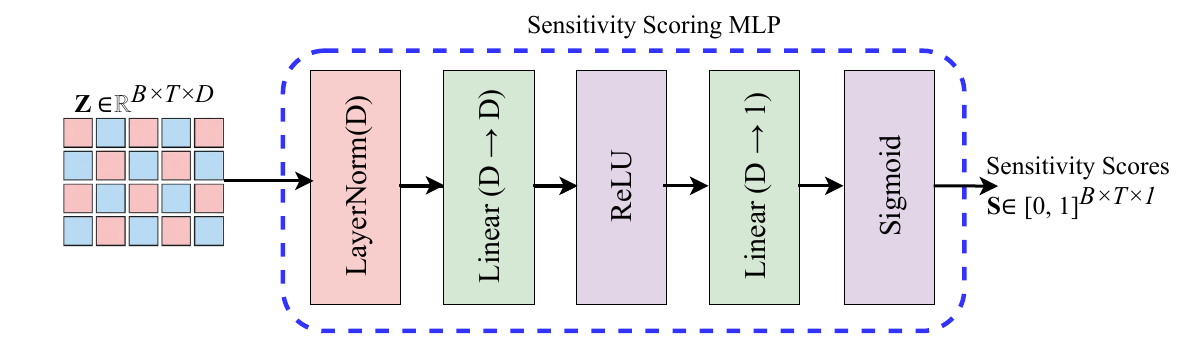}
  \caption{Sensitivity scoring MLP. The module estimates token-wise sensitivity scores $\mathbf{S}\in[0,1]$ from $\mathbf{Z}$.}
  \label{fig:stf}
\end{figure}

 \begin{figure}[!t]
  \centering
  \includegraphics[width=0.85\linewidth]{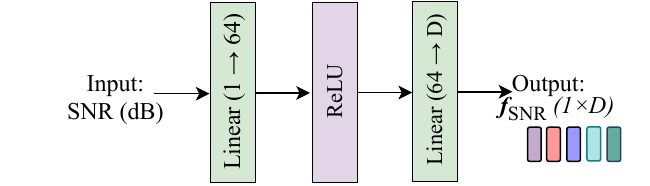}  
  \caption{SNR embedding module.}
  \label{fig:snr}
\end{figure}

\subsection{Sensitivity-based Token Filtering (STF)}\label{sec:stf}
After encoding, different latent tokens may contribute unequally to reconstruction distortion under quantization and channel perturbation. STF assigns each token a sensitivity
 score using a lightweight MLP, as shown in Fig.~\ref{fig:stf}. Given encoder tokens $\mathbf{Z}\in\mathbb{R}^{B\times T\times D}$, the scores are computed as
\begin{equation}
\mathbf{S}=\sigma\!\left(\text{Linear}_{D\rightarrow 1}\!\left(\text{ReLU}\!\left(\text{Linear}_{D\rightarrow D}\!\left(\text{LayerNorm}(\mathbf{Z})\right)\right)\right)\right),
\end{equation}
where $\mathbf{S}\in\mathbb{R}^{B\times T\times 1}$ and $\sigma(\cdot)$ is the sigmoid function. Each score reflects the token's learned sensitivity to reconstruction distortion under quantization and channel perturbation.
During training, the scores act as soft weights applied to the tokens:
\begin{equation}
\tilde{\mathbf{Z}}=\mathbf{Z}\odot\mathbf{S},
\end{equation}
where $\odot$ denotes element-wise multiplication. We further refine the weighted tokens using a lightweight MLP:
\begin{equation}
\mathbf{Z}'=\text{Linear}_{D\rightarrow D}\!\left(\text{ReLU}\!\left(\text{Linear}_{D\rightarrow D}\!\left(\text{LayerNorm}(\tilde{\mathbf{Z}})\right)\right)\right),
\end{equation}
and forward $\mathbf{Z}'$ to the quantization and channel transmission blocks. 

\subsection{Quantization}\label{sec:quantization}
After sensitivity-based token filtering, the filtered tokens are mapped to a discrete representation for transmission. To enable end-to-end training, we use uniform integer quantization with a learnable scaling factor. We introduce a training-only symbol-usage penalty to stabilize the use of discrete codes. Tokens are normalized per sample:
\begin{equation}
\hat{\mathbf{Z}} = \frac{\mathbf{Z}' - \mu}{\sigma_z + \epsilon},
\end{equation}
where $\mu$ and $\sigma_z$ are the mean and standard deviation of $\mathbf{Z}'$ across the feature dimension $D$, and $\epsilon = 10^{-6}$. A learnable global scaling factor $\alpha$ is then applied:
\begin{equation}
\mathbf{Z}_{q} = \text{Round}\!\left(\alpha \cdot \hat{\mathbf{Z}}\right),
\end{equation}
where $\text{Round}(\cdot)$ denotes element-wise rounding and $\alpha$ is constrained to be positive via a softplus transform. During training, we use the straight-through estimator (STE) to approximate gradients through rounding:
\begin{equation}
\frac{\partial \mathbf{Z}_{q}}{\partial \hat{\mathbf{Z}}} \approx 1.
\end{equation}
The symbol-usage statistics used by the penalty are computed only during training.

\subsection{SNR-Aware Embedding Injection}\label{sec:snr}
We model wireless transmission by applying channel perturbations to the quantized tokens. For a target SNR (dB), with unit-power normalization, the noise variance is set as
\begin{equation}
\sigma_{\text{noise}}^{2} = 10^{-\text{SNR}/10}.
\end{equation}
Under an AWGN channel, the received tokens are
\begin{equation}
\mathbf{Y}_{\text{AWGN}} = \mathbf{Z}_{q} + \mathbf{n}, \quad \mathbf{n} \sim \mathcal{N}(0, \sigma_{\text{noise}}^{2}).
\end{equation}
Under a Rayleigh fading channel, we use
\begin{equation}
\mathbf{Y}_{\text{Rayleigh}} = h \cdot \mathbf{Z}_{q} + \mathbf{n}, \quad h \sim \text{Rayleigh}(\theta),
\end{equation}
where $h$ is sampled independently per test sample and applied to all transmitted tokens (block fading),
and $\theta = \tfrac{\sqrt{2}}{2}$ \cite{simon2014fading}. We denote the received tokens generically as $\mathbf{Y}$.

To adapt decoding to channel conditions, we inject a lightweight SNR embedding into the received tokens as illustrated in Fig.~\ref{fig:snr}. In our evaluation, the decoder is provided with a receiver-side SNR estimate as side information.
The estimate is mapped through a two-layer MLP:
\begin{equation}
\mathbf{f}_{\text{SNR}} = \text{Linear}_{64 \rightarrow D}\!\left(\text{ReLU}\!\left(\text{Linear}_{1 \rightarrow 64}(\text{SNR})\right)\right),
\end{equation}
The resulting embedding $\mathbf{f}_{\mathrm{SNR}}\in\mathbb{R}^{1\times 1\times D}$ is broadcasted over the batch and token axes and added to $\mathbf{Y}$:
\begin{equation}
\mathbf{Y}' = \mathbf{Y} + \mathbf{f}_{\text{SNR}}.
\end{equation}
This additive injection conditions the decoder features on the estimated channel quality while keeping the module lightweight. 
\begin{figure*}[!t]
  \centering
  \includegraphics[width=0.70\textwidth]{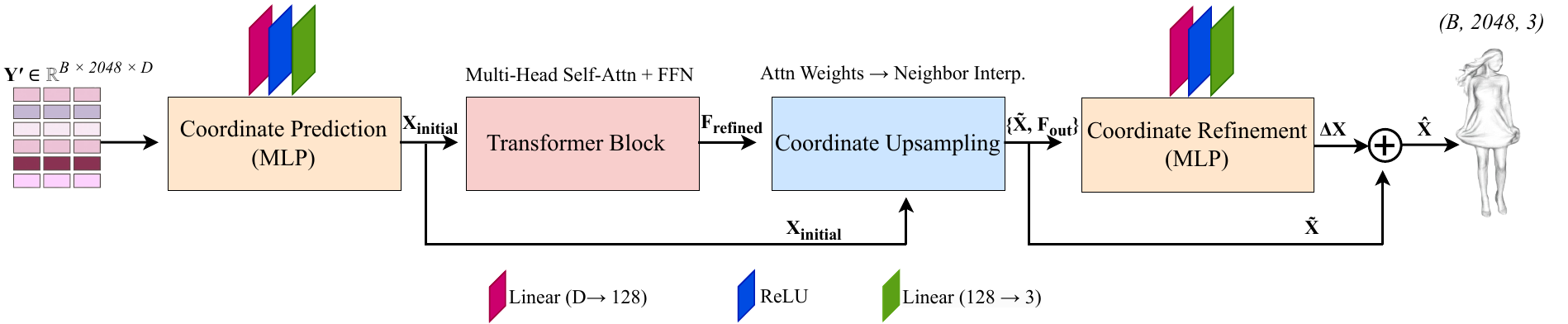}
  \caption{Detailed architecture of the refinement-based decoder for 3D point cloud reconstruction.}
  \label{fig:decoder}
\end{figure*}

\begin{figure}[!t]
  \centering
  \includegraphics[width=0.92\linewidth]{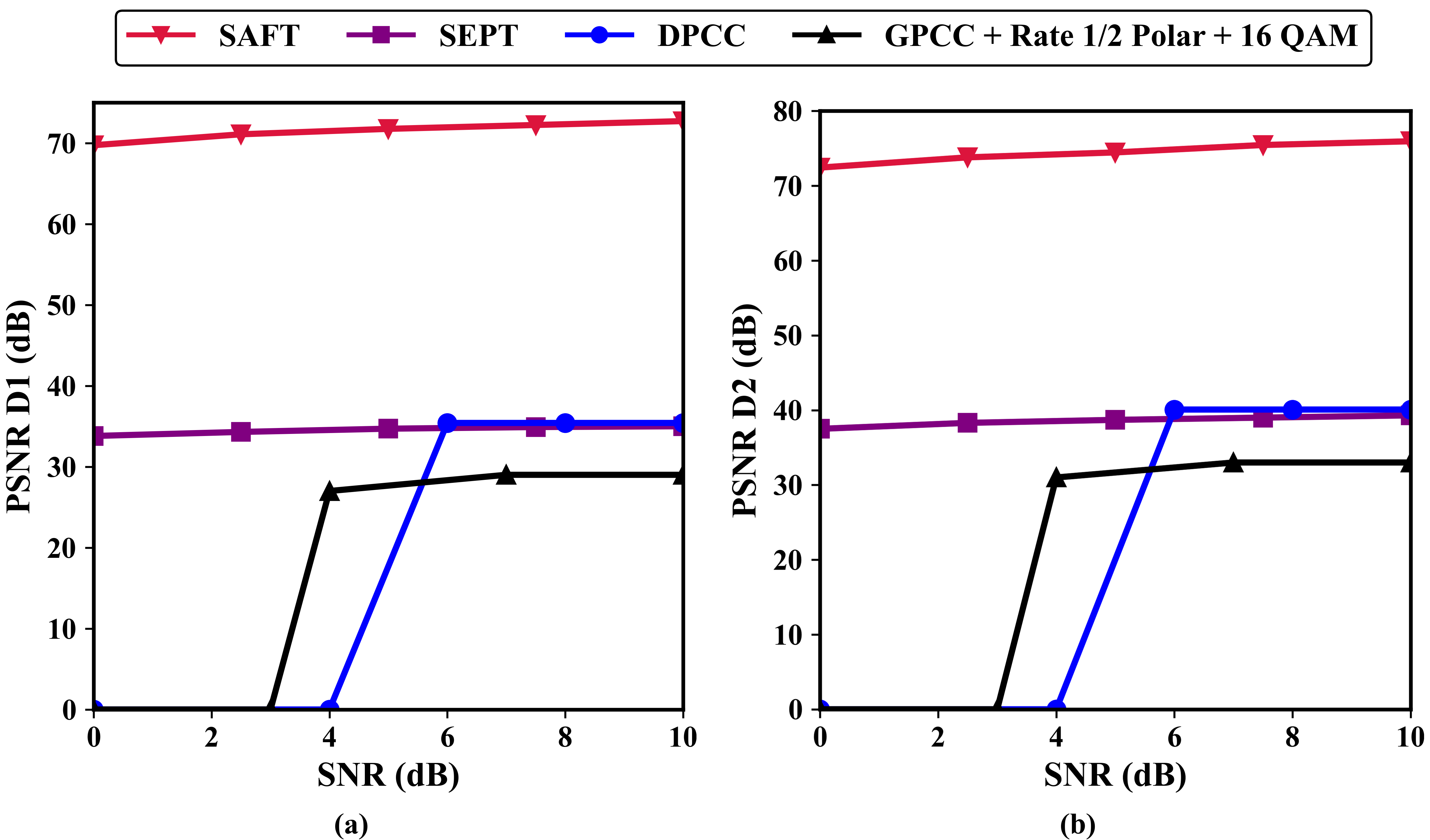}
  \caption{Performance comparison on the ShapeNet dataset across SNR levels.
  (a) D1 PSNR; (b) D2 PSNR.}
  \label{fig:shapenet_perf}
  \vspace{-2mm}
\end{figure}

\subsection{Decoder Architecture}\label{sec:decoder}
As shown in Fig.~\ref{fig:decoder}, the decoder maps the received tokens after SNR embedding injection $\mathbf{Y}'$ to a reconstructed point cloud. A linear projection first produces a coarse coordinate estimate:
\begin{equation}
\mathbf{X}_{\text{initial}} = \text{Linear}_{D \rightarrow 3}(\mathbf{Y}').
\end{equation}

A lightweight transformer block then refines token features by modeling dependencies across tokens conditioned on the coarse coordinates:
\begin{equation}
\mathbf{F}_{\text{refined}} = \text{TransformerBlock}(\mathbf{Y}', \mathbf{X}_{\text{initial}}).
\end{equation}

Next, an attention-guided upsampling module increases spatial resolution by generating an $N$-point set and corresponding features:
\begin{equation}
\{\tilde{\mathbf{X}}, \mathbf{F}_{\text{out}}\} = \text{Upsample}(\mathbf{X}_{\text{initial}}, \mathbf{F}_{\text{refined}}).
\end{equation}
Finally, an MLP head predicts a residual coordinate correction:
\begin{equation}
\hat{\mathbf{X}} = \tilde{\mathbf{X}} + 
\text{Linear}_{D \rightarrow 3}\!\left(
\text{ReLU}\!\left(\text{Linear}_{D \rightarrow 128}(\mathbf{F}_{\text{out}})\right)
\right).
\end{equation}

This final residual MLP head predicts a per-point coordinate correction that helps mitigate local distortions introduced by quantization and channel perturbations.
The final reconstructed point cloud $\hat{\mathbf{X}}$ is evaluated against the ground truth $\mathbf{X}$ using standard geometric distance metrics, completing the end-to-end recovery process.

\subsection{Regularization and Loss Functions}\label{sec:loss}
We train SAFT using a multi-term objective that combines reconstruction loss with auxiliary penalty terms:
\begin{equation}
\mathcal{L}_{\text{total}} = \mathcal{L}_{\text{CD}}
+ \lambda_{\text{sym}} \mathcal{L}_{\text{sym}}
+ \lambda_{\text{sparsity}} \mathcal{L}_{\text{sparsity}}
+ \lambda_{\text{diversity}} \mathcal{L}_{\text{diversity}}.
\end{equation}

The reconstruction term is the chamfer distance (CD)~\cite{ref26_cvpr_psgn} between the ground-truth point cloud $\mathbf{X}$ and the reconstruction $\hat{\mathbf{X}}$:
\begin{equation}
\mathcal{L}_{\text{CD}} = 
\frac{1}{|\mathbf{X}|} \sum_{\mathbf{x} \in \mathbf{X}} 
\min_{\hat{\mathbf{x}} \in \hat{\mathbf{X}}} \|\mathbf{x} - \hat{\mathbf{x}}\|_{2}^{2} 
+ \frac{1}{|\hat{\mathbf{X}}|} \sum_{\hat{\mathbf{x}} \in \hat{\mathbf{X}}} 
\min_{\mathbf{x} \in \mathbf{X}} \|\hat{\mathbf{x}} - \mathbf{x}\|_{2}^{2}.
\end{equation}

To  discourage degenerate utilization in the quantized latent space, $\mathcal{L}_{\text{sym}}$ is computed only during training using histogram-spread statistics over  $\mathbf{Z}_q = Q(\mathbf{Z}')$.

To regularize token scoring, we apply a sparsity penalty on the STF scores:
\begin{equation}
\mathcal{L}_{\text{sparsity}} = \frac{1}{BT}\sum_{b=1}^{B}\sum_{i=1}^{T}\max(S_{b,i}-\tau,0), \quad \tau=0.1.
\end{equation}
To mitigate redundancy among tokens, we introduce a decorrelation-based diversity term~\cite{ref29_icml_barlow_twins}. This term discourages token collapse by penalizing inter-token correlation; its weight is tuned to avoid disrupting geometric continuity:
\begin{equation}
\mathcal{L}_{\text{diversity}} = \frac{1}{B}\sum_{b=1}^{B}\frac{1}{T(T-1)}\sum_{i\neq j}
\left(\frac{1}{D}\,
\mathbf{Z}^{(i)}_{q,b}\cdot \mathbf{Z}^{(j)}_{q,b}\right)^{2},
\end{equation}
where $\mathbf{Z}^{(i)}_{q,b}\in\mathbb{R}^{D}$ denotes the $i$-th normalized token of the $b$-th sample.

\newcommand{\na}{N/A}  

\begin{table*}[t]
\caption{D1 and D2 PSNR on \textit{8iVFB} under AWGN and Rayleigh at SNRs of 0 and 10 dB.}
\label{tab:8ivfb_psnr}
\centering
\scriptsize
\setlength{\tabcolsep}{2.8pt}
\renewcommand{\arraystretch}{1.12}
\begin{tabular}{|l|l|
                cc|cc|
                cc|cc|
                cc|cc|
                cc|cc|
                cc|cc|}
\hline
\multicolumn{2}{|c|}{\textbf{Method}} &
\multicolumn{4}{c|}{\textbf{SEPT}} &
\multicolumn{4}{c|}{\textbf{G-PCC (octree)+LDPC}} &
\multicolumn{4}{c|}{\textbf{V-PCC+LDPC}} &
\multicolumn{4}{c|}{\textbf{PCST}} &
\multicolumn{4}{c|}{\textbf{SAFT}} \\
\cline{3-22}
\multicolumn{2}{|c|}{} &
\multicolumn{2}{c|}{\textbf{AWGN}} &
\multicolumn{2}{c|}{\textbf{Rayleigh}} &
\multicolumn{2}{c|}{\textbf{AWGN}} &
\multicolumn{2}{c|}{\textbf{Rayleigh}} &
\multicolumn{2}{c|}{\textbf{AWGN}} &
\multicolumn{2}{c|}{\textbf{Rayleigh}} &
\multicolumn{2}{c|}{\textbf{AWGN}} &
\multicolumn{2}{c|}{\textbf{Rayleigh}} &
\multicolumn{2}{c|}{\textbf{AWGN}} &
\multicolumn{2}{c|}{\textbf{Rayleigh}} \\
\cline{3-22}
\multicolumn{2}{|c|}{} &
\textbf{0 dB} & \textbf{10 dB} &
\textbf{0 dB} & \textbf{10 dB} &
\textbf{0 dB} & \textbf{10 dB} &
\textbf{0 dB} & \textbf{10 dB} &
\textbf{0 dB}& \textbf{10 dB} &
\textbf{0 dB} & \textbf{10 dB} &
\textbf{0 dB} & \textbf{10 dB} &
\textbf{0 dB} & \textbf{10 dB} &
\textbf{0 dB} & \textbf{10 dB} &
\textbf{0 dB} & \textbf{10 dB} \\
\hline

\multicolumn{22}{|c|}{\textbf{Soldier}} \\
\hline
\multirow{3}{*}{\textbf{Metrics}}
& D1$\uparrow$
  & 29.34 & 30.37 & 27.83 & 28.43
  & --    & 70.75  & --    & --
  & --    & \textbf{71.84}  & --    & 67.81
  & \na   & 71.42  & \na   & \textbf{70.72}
  & \textbf{69.82} & 71.78  & \textbf{64.87} & 69.81 \\
& D2$\uparrow$
  & 30.54 & 31.91 & 28.86 & 29.59
  & --    & 75.55  & --    & --
  & --    & 74.99  & --    & 69.76
  & \na   & 74.82  & \na   & 74.06
  & \textbf{74.58} & \textbf{76.50}  & \textbf{69.94} & \textbf{74.43} \\
& Avg$\uparrow$
  & 29.94 & 31.14 & 28.35 & 29.01
  & --    & 73.15  & --    & --
  & --    & 73.42  & --    & 68.79
  & \na   & 73.12  & \na   & \textbf{72.39}
  & \textbf{72.20} & \textbf{74.14}  & \textbf{67.41} & 72.12 \\
\hline

\multicolumn{22}{|c|}{\textbf{Loot}} \\
\hline
\multirow{3}{*}{\textbf{Metrics}}
& D1$\uparrow$
  & 27.93 & 31.05 & 28.31 & 29.69
  & --    & 70.71  & --    & --
  & --    & 72.25  & --    & 68.42
  & \na   & \textbf{72.51}  & \na   & \textbf{71.25}
  & \textbf{69.66} & 71.50  & \textbf{64.46} & 69.82 \\
& D2$\uparrow$
  & 30.28 & 32.51 & 30.29 & 30.85
  & --    & 75.45  & --    & --
  & --    & 75.39 & --    & 70.42
  & \na   & 76.09  & \na   & 74.61
  & \textbf{74.71} & \textbf{76.71}  & \textbf{69.32} & \textbf{74.65} \\
& Avg$\uparrow$
  & 29.11 & 31.78 & 29.30 & 30.27
  & --    & 73.08  & --    & --
  & --    & 73.82  & --    & 69.42
  & \na   & \textbf{74.30}  & \na   & \textbf{72.93}
  & \textbf{72.19} & 74.11  & \textbf{66.89} & 72.24 \\
\hline

\multicolumn{22}{|c|}{\textbf{Red and black}} \\
\hline
\multirow{3}{*}{\textbf{Metrics}}
& D1$\uparrow$
  & 30.14 & 32.69 & 29.69 & 31.38
  & --    & 69.99  & --    & --
  & --    & 71.64 & --    & 67.73
  & \na   & 70.62  & \na   & 69.92
  & \textbf{70.69} & \textbf{72.92}  & \textbf{65.33} & \textbf{70.05} \\
& D2$\uparrow$
  & 33.57 & 34.86 & 30.73 & 32.62
  & --    & 74.86  & --    & --
  & --    & 74.71  & --    & 69.70
  & \na   & 74.41  & \na   & 73.63
  & \textbf{75.77} & \textbf{77.05}  & \textbf{69.78} & \textbf{74.29} \\
& Avg$\uparrow$
  & 31.85 & 33.78 & 30.21 & 32.00
  & --    & 72.42  & --    & --
  & --    & 73.18 & --    & 68.72
  & \na   & 72.52  & \na   & 71.78
  & \textbf{73.23} & \textbf{74.99}  & \textbf{67.55} & \textbf{72.17} \\
\hline
\end{tabular}

{\footnotesize\itshape
\vspace{4pt}
\textit{Note.} Entries are D1/D2 PSNR. “--” indicates decoding failure at that SNR; “\na” = result unavailable (PCST results at SNR = 0 dB are not available).}
\end{table*}

\begin{figure}[!t]
\centering
\includegraphics[width=\linewidth]{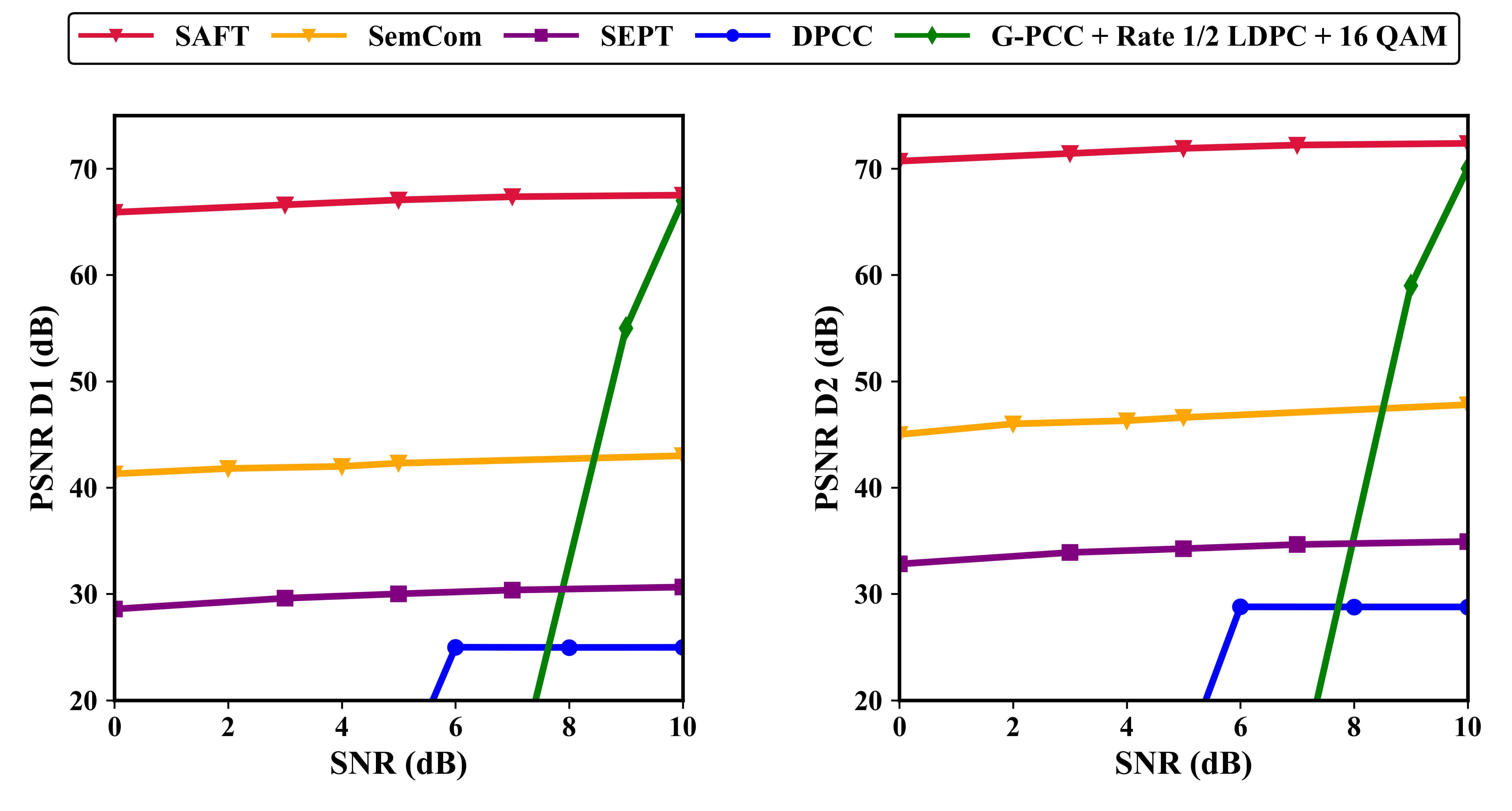} 
\caption{Performance comparison on the ModelNet40 dataset across SNR levels. (a) D1 PSNR; (b) D2 PSNR.}
\label{fig:modelnet40_perf}
\end{figure}

\begin{figure*}[!t]
\centering
\includegraphics[width=0.85\textwidth]{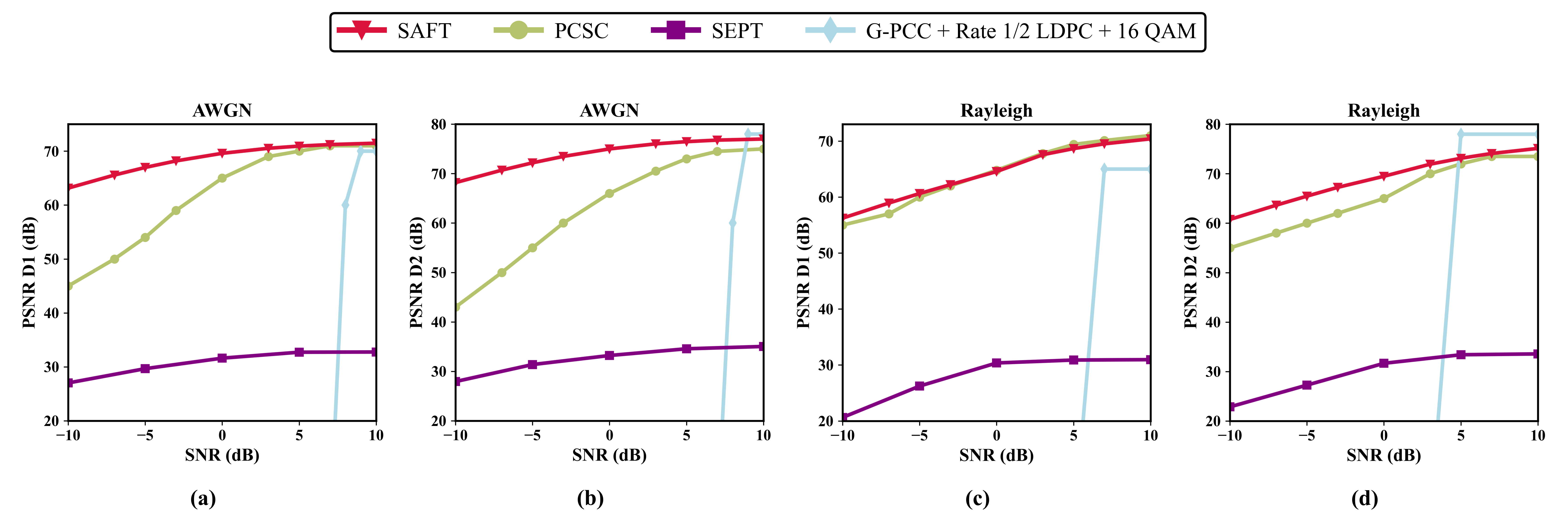} 
\caption{Robust reconstruction on \textit{Longdress} under AWGN and Rayleigh channels. (a) D1 (AWGN); (b) D2 (AWGN); (c) D1 (Rayleigh); (d) D2 (Rayleigh).}
\label{fig:longdress_robust}
\end{figure*}
\begin{figure}[!t]
\centering
\includegraphics[width=\columnwidth]{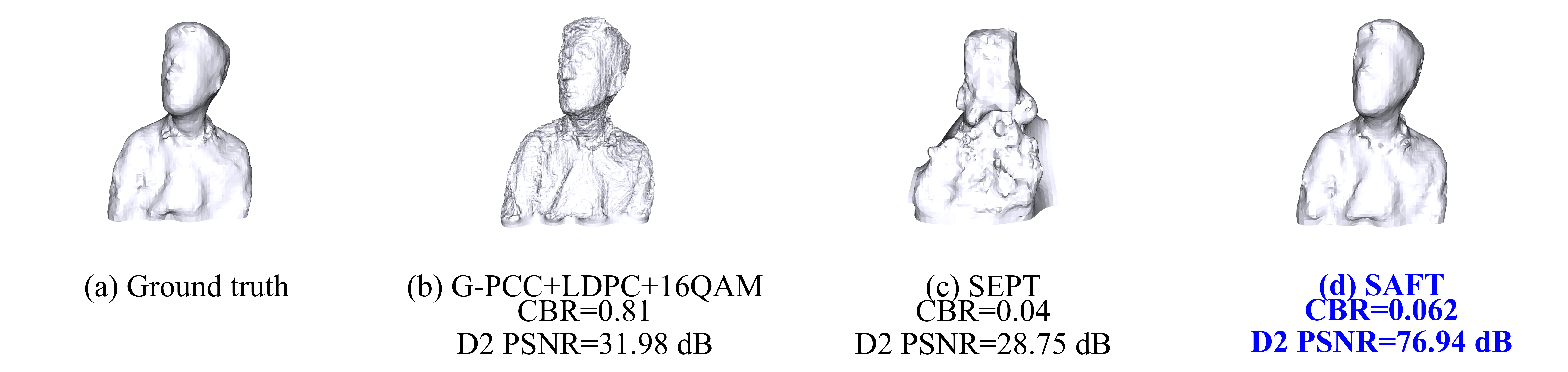} 
\caption{Visual comparison of the \textit{Andrew} under an AWGN channel with SNR = 10 dB.}
\label{fig:andrew_vis}
\end{figure}

\begin{figure}[!t]
  \centering
  \includegraphics[width=\linewidth]{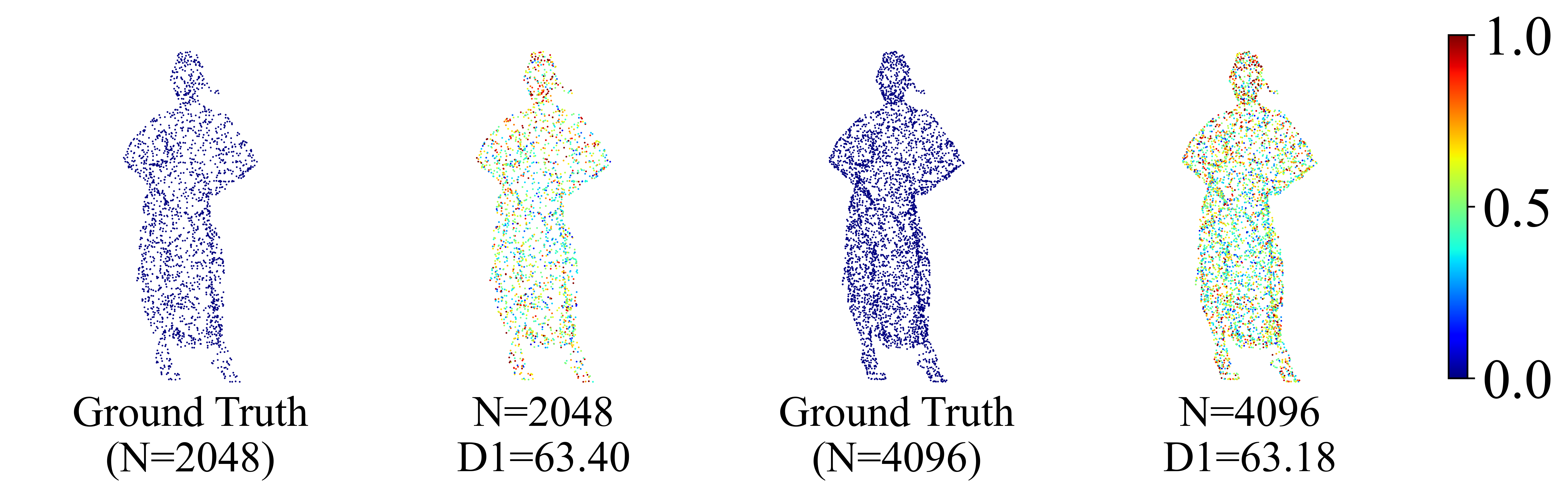}
  \caption{Input-density robustness at $\mathrm{SNR}=-10$~dB: $N{=}2048$ vs.\ $N{=}4096$. Colors show normalized pointwise error.}
  \label{fig:density_robust}
\end{figure}

\begin{figure}[!t]
\centering
\vspace{-2mm}
\includegraphics[width=0.68\linewidth]{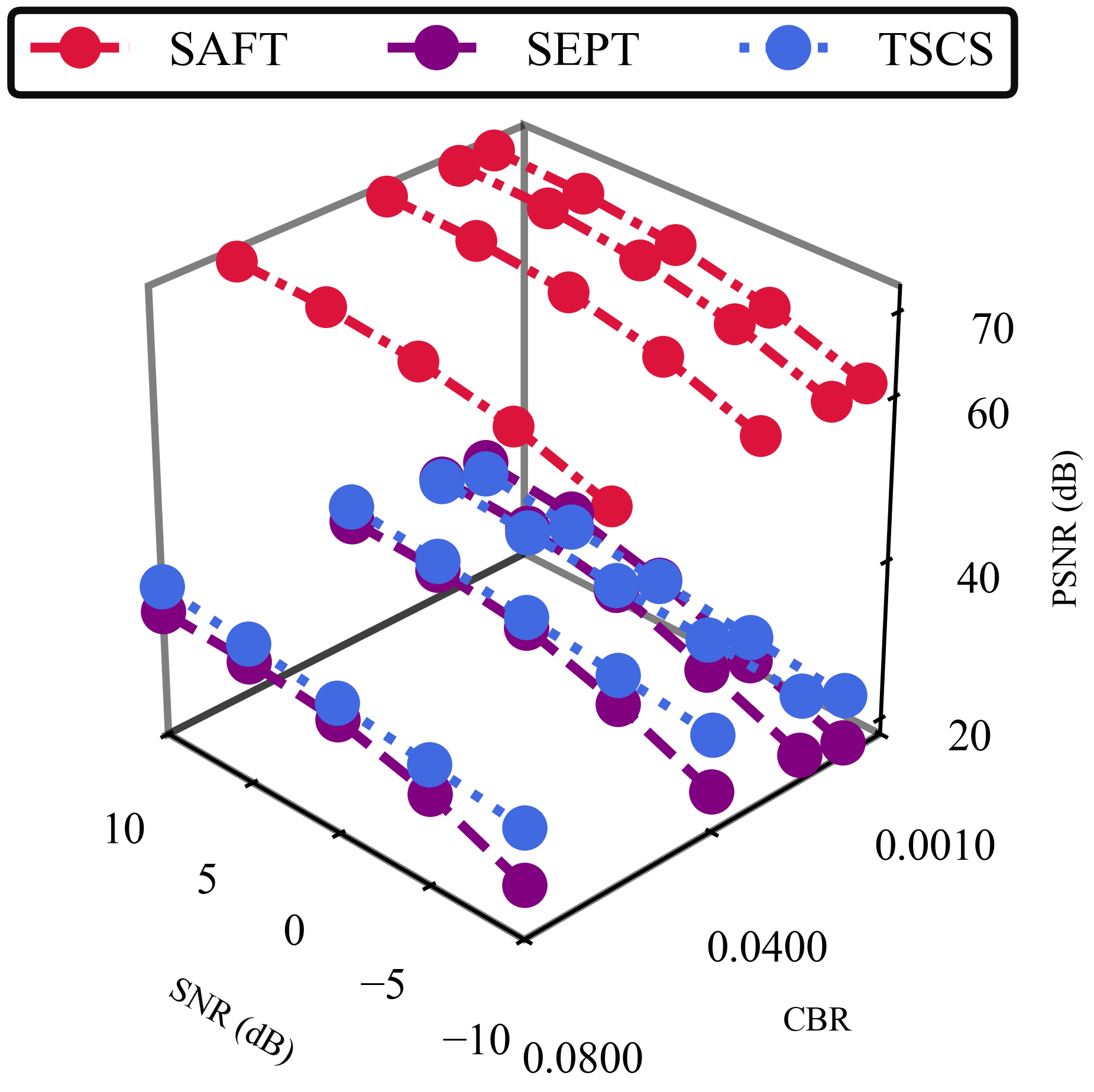}
\vspace{-2mm}
\caption{D1 PSNR across CBR at different SNR levels.}
\label{fig:cbr_d1}
\vspace{-2mm}
\end{figure}

\begin{figure}[!t]
\centering
\includegraphics[width=\linewidth]{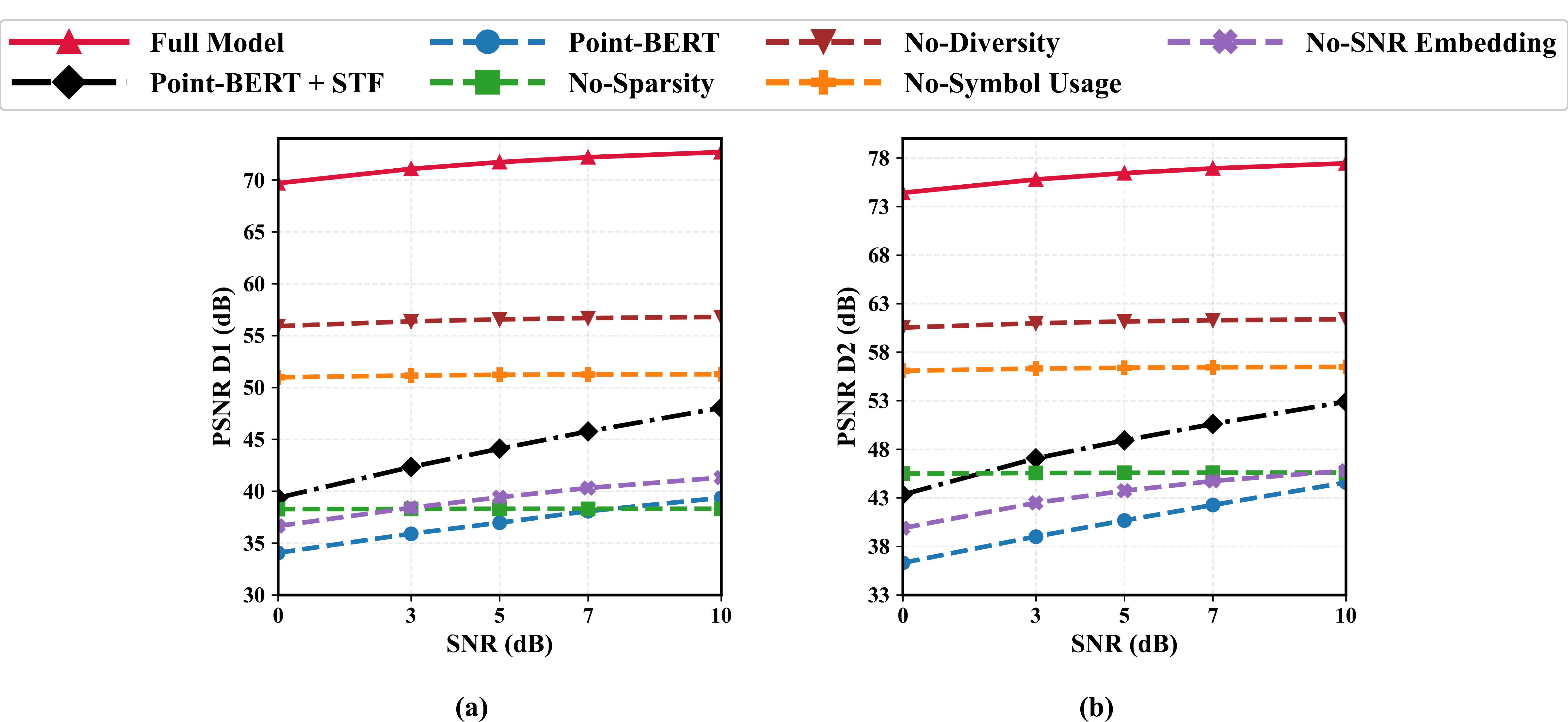}  
\caption{Ablation study of the SAFT architecture under varying SNRs. (a) D1 PSNR; (b) D2 PSNR.}
\label{fig:ablation_saft}
\end{figure}
\begin{figure}[!t]
  \centering
  \includegraphics[width=\linewidth]{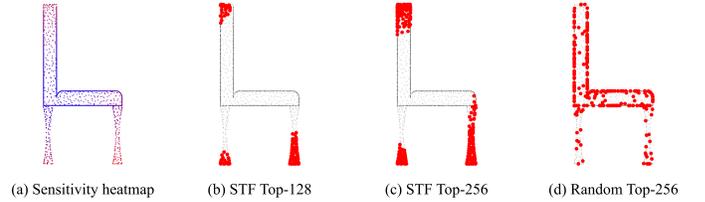}
  \caption{ STF sensitivity visualization and token selection examples. (a) Learned sensitivity heatmap. (b) STF Top-$128$. (c) STF Top-$256$. (d) Random Top-$256$ selection.}
  \label{fig:saliency_vis}
\end{figure}

\begin{table}[t]
\centering
\caption{Sensitivity to SNR estimate offset at SNR $=-10$~dB (D1 PSNR).}
\label{tab:snr_mismatch}
\begin{tabular}{|c|c|}
\hline
\textbf{Offset $\Delta$ (dB)} & \textbf{D1 PSNR (dB)} \\
\hline
$-2$ & 60.29 \\
$0$  & 61.77 \\
$+2$ & 63.08 \\
\hline
\end{tabular}%

\end{table}
\begin{table}[t]
\centering
\caption{Decoder-step ablation under stress-test setting (SNR $=-20$ dB).}
\label{tab:decoder_ablation_snr20}
\begin{tabular}{lcc}
\hline
Setting & D1 PSNR & D2 PSNR \\
\hline
Full & \textbf{53.33} & \textbf{58.08} \\
--Residual MLP & 31.36 & 33.08 \\
--Upsample & 31.26 & 32.96 \\
--Transformer & 53.32 & 58.08 \\
Coarse-only & 31.27 & 32.96 \\
\hline
\end{tabular}%

\end{table}

\begin{table}[t]
\centering
\caption{Model complexity comparison.}
\label{tab:model_complexity}
\begin{tabular}{|l|c|c|c|}
\hline
\textbf{Method} & \textbf{Params (M)} & \textbf{FLOPs (G)} & \textbf{Latency (ms/pc)} \\
\hline
SAFT & 4.17 & 15.47 & 14.51\\
TSCS & 9.87 & 13.05 & 20.10\\
SEPT & 10.22 & 12.36 & 21.51\\
\hline
\end{tabular}
\end{table}
\section{Experimental Results and Analyses}\label{sec:experiments}
This section evaluates the proposed SAFT under different
wireless channel conditions. The focus is on three aspects: reconstruction
quality, CBR efficiency, and noise robustness.

\subsection{Experimental Setup}\label{sec:setup}
SAFT was implemented in PyTorch (v1.13.0) and trained on an NVIDIA RTX 4090 GPU. We trained and evaluated on ShapeNet \cite{ref31_arxiv_shapenet} using a randomized 90/10 train--test split, and on ModelNet40 \cite{ref32_cvpr_3dshapenets} using the official split. All point clouds were downsampled to $N{=}2048$ points via farthest point sampling (FPS) for controlled comparison across methods. For real-scene evaluation, we additionally tested on four representative full-body point clouds from 8iVFB under both AWGN and Rayleigh channels. We used Adam with batch size 16, learning rate $10^{-3}$, weight decay $10^{-4}$, and early stopping (max 100 epochs) during training. We set $\lambda_{\text{sym}}{=}0.5$ and fix $\lambda_{\text{sparsity}}{=}\lambda_{\text{diversity}}{=}1.0$ in all experiments empirically.

We compared SAFT with conventional and learning-based baselines. For separate source--channel coding (SSCC), we used MPEG G-PCC (TMC13v28) and V-PCC (TMC2v25), whose bitstreams were protected by rate $1/2$ LDPC coding and transmitted over AWGN and Rayleigh channels under matched channel conditions (reported as $E_b/N_0$ per coded bit for SSCC). Specifically, G-PCC used an IEEE 802.11 LDPC code ($n{=}648,k{=}324$) with 16-QAM and 20 decoding iterations, while V-PCC used an LDPC code ($n{=}1200,k{=}601$) with BPSK and 80 decoding iterations. For Rayleigh fading, block fading was applied per LDPC codeword with the fading coefficient estimated and equalized before LDPC decoding. Learning-based baselines include DPCC \cite{ref34_apccpa_transformer_upsampling_pcc}, SemCom \cite{ref30_globecom_semcom_pc}, SEPT \cite{ref35_spawc_wireless_pc}, PCST \cite{ref8_icassp_pcst}, TSCS \cite{ref36_iwrfat_tscs_pointcloud}, and PCSC \cite{ref9_tvt_semcom_pointcloud}. 
We computed D1-PSNR (point-to-point) and D2-PSNR (point-to-plane) using the MPEG point-cloud distortion evaluation procedure \cite{refXX_mpeg_gpcc_ctc_n00722} with identical metric settings for all reproduced methods. 
For SemCom, PCST, and PCSC, results are cited from the original papers when implementations are unavailable and are included for reference.
All point clouds are normalized to a common bounding box before evaluation, and the same normalization is applied to all methods when computing D1 and D2 PSNR, which affects the PSNR numerical scale.

\subsection{Reconstruction Quality Evaluation}\label{sec:quality}
To evaluate reconstruction performance, we report D1 and D2 PSNR on the test split of the ShapeNet dataset.
As shown in Fig.~\ref{fig:shapenet_perf}, SAFT maintains strong reconstruction quality across all SNR conditions.
Compared with competing methods, SAFT achieves higher D1 and D2 PSNR than SEPT \cite{ref35_spawc_wireless_pc}, DPCC \cite{ref34_apccpa_transformer_upsampling_pcc}, and the SSCC.
The performance margin increases at low SNRs, where conventional codecs and prior learned baselines degrade more rapidly.

We also evaluate on ModelNet40 using a model trained on the ModelNet40 training split and tested on the official test split.
Fig.~\ref{fig:modelnet40_perf} shows D1 and D2 PSNR across SNR levels.
SAFT achieves strong performance across all SNRs while maintaining stable behavior under degraded channel conditions.
Absolute PSNR values vary across datasets due to differences in geometric complexity, while the relative trends across methods remain consistent. For the SSCC baseline (G-PCC + LDPC + 16-QAM), PSNR drops sharply at low SNR, consistent with the threshold behavior commonly observed in channel-coded pipelines.

Fig.~\ref{fig:longdress_robust} reports results on \textit{Longdress} under AWGN and Rayleigh fading from $-10$~dB to 10~dB. Table~\ref{tab:8ivfb_psnr} presents results on additional 8iVFB sequences (\textit{Soldier}, \textit{Redandblack}, and \textit{Loot}) under AWGN and Rayleigh at 0~dB and 10~dB. We can see that SAFT maintains strong performance across all cases, demonstrating robustness to channel degradation.
Fig.~\ref{fig:andrew_vis} shows a visual comparison of reconstructed point clouds from G-PCC, SEPT, and the proposed method, alongside the ground truth. The proposed method preserves shape and structure more effectively while achieving a better rate--distortion balance.
We further test SAFT with $N{=}4096$ input points to assess scalability. The D1 PSNR as shown in Fig.~\ref{fig:density_robust} remains nearly unchanged (63.40~dB vs.\ 63.18~dB), confirming robustness under increased point density.

\subsection{Impact of Channel Bandwidth Ratio (CBR) on Reconstruction Quality}\label{sec:cbr}
Fig.~\ref{fig:cbr_d1} reports D1 PSNR versus CBR across SNRs from $-10$~dB to $10$~dB. We can see that SAFT outperforms SEPT and TSCS, achieving much better quality at a lower CBR, which is computed from the transmitted payload: After STF, we send the top-$K$ token features ($K \times D$ quantized elements) normalized by $3N$, yielding $\mathrm{CBR}=\frac{KD}{3N}$, which is a deterministic payload ratio controlled by $K$. With a fixed $b$-bit scalar quantizer, the bpp-equivalent payload is $R_{\mathrm{bpp}}=\frac{KDb}{N}$. In the reported experiments, CBR is fixed by the chosen operating setting through the selected token budget $K$, rather than being adapted on a per-sample basis.

\subsection{Ablation Study}\label{sec:ablation}

Fig.~\ref{fig:ablation_saft} reports ablations under the same evaluation protocol with matched evaluation settings. Comparing the backbone with and without STF (Point-BERT vs.\ Point-BERT+STF) shows a consistent gain, which is consistent with the benefit of token-wise importance weighting for robust transmission. Removing SNR embedding (No-SNR Embedding) reduces PSNR across SNRs, reflecting weaker adaptation to channel quality variations. Removing the training-only symbol-usage penalty (No-Symbol Usage) also degrades performance, consistent with the need to stabilize discrete code utilization. Dropping sparsity (No-Sparsity) or diversity (No-Diversity) leads to smaller but consistent degradations. Fig.~\ref{fig:saliency_vis} further provides qualitative evidence of sensitivity maps and Top-$K$ token selection compared with random Top-$K$ selection.
Table~\ref{tab:snr_mismatch} evaluates sensitivity to imperfect receiver-side SNR information by perturbing the decoder-side SNR input (receiver estimate) as $\mathrm{SNR}_{\text{in}}=\mathrm{SNR}+\Delta$ during testing D1 PSNR at SNR $=-10$~dB. The results vary by 2.79 dB over $\Delta \in [-2, +2]$, indicating moderate sensitivity without failure at the receiver.
We further examine the decoder design by ablating the reconstruction stages as shown in Table~\ref{tab:decoder_ablation_snr20}. The upsample stage and the final residual MLP head contribute most to geometric fidelity, while transformer refinement yields a negligible change under this setting, serving as a global-context stabilizer.
\subsection{Computational Complexity and Runtime Considerations}\label{sec:complexity}
Table~\ref{tab:model_complexity} summarizes model complexity and runtime. SAFT uses 4.17M parameters, which are smaller than TSCS (8.65M) and SEPT (10.22M). THOP profiling reports 15.47 GFLOPs for SAFT, reflecting additional computation from token filtering and refinement. We also report measured inference latency (ms/pc) on an RTX 4090 GPU under an evaluation setting (batch size $=16$) after warm-up.
\section{Conclusion}\label{sec:conclusion}
This paper presented SAFT, a learned transmission framework for wireless delivery of 3D point clouds. SAFT combines token-wise  sensitivity weighting with SNR-conditioned decoding to improve reconstruction quality under channel perturbations. Experiments on multiple datasets show that SAFT achieves higher D1/D2 PSNR than conventional and learning-based baselines, with gains more pronounced under degraded channel conditions. Future work will study task-level utility using downstream evaluations such as segmentation or classification.

\bibliographystyle{IEEEtran}
\bibliography{references}
\begin{IEEEbiography}[{\includegraphics[width=1in,height=1.25in,clip,keepaspectratio]{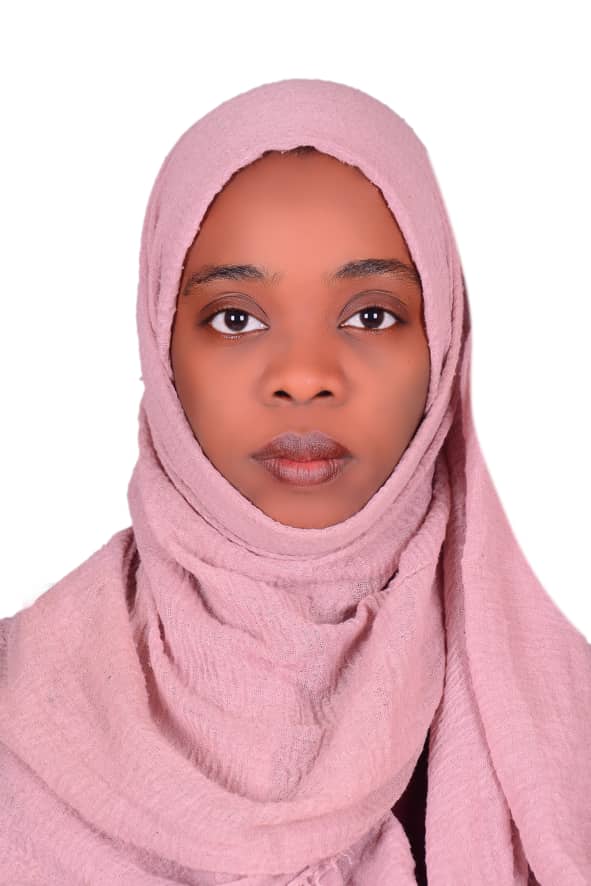}}]{Huda Adam Sirag Mekki}
received the B.Sc. degree in Computer Engineering in 2011 and the M.Sc. degree in Computer Engineering and Networks in 2015, both from the University of Gezira, Sudan. Since 2015, she has been a Lecturer with the Department of CE. She is currently pursuing the Ph.D. degree with the School of Control Science and Engineering, Shandong University, Jinan, China. Her research include semantic communication for 3D point clouds.
\end{IEEEbiography}

\begin{IEEEbiography}[{\includegraphics[width=1in,height=1.25in,clip,keepaspectratio]{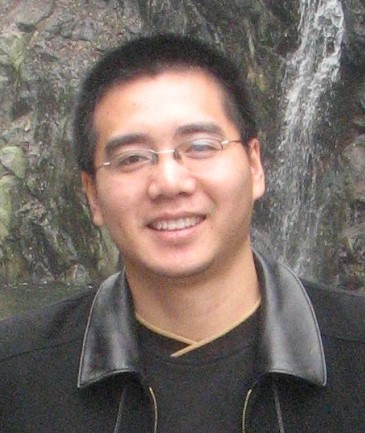}}]{Hui Yuan}
(Senior Member, IEEE) received the B.E. and Ph.D. degrees in telecommunication engineering from Xidian University, Xi’an, China, in 2006 and 2011, respectively. In April 2011, he joined Shandong University, Jinan, China, as a Lecturer (April 2011–December 2014), an Associate Professor (January 2015–August 2016), and a Professor (September 2016). From January 2013 to December 2014 and from November 2017 to February 2018, he was a Postdoctoral Fellow (Granted by the Hong Kong Scholar Project) and a Research Fellow, respectively, with the Department of Computer Science, City University of Hong Kong. From November 2020 to November 2021, he was a Marie Curie Fellow (Granted by the Marie Marie Sk\l{}odowska-Curie Actions Individual Fellowship under Horizon2020 Europe) with the School of Engineering and Sustainable Development, De Montfort University, Leicester, U.K. From October 2021 to November 2021, he was also a Visiting Researcher (secondment of the Marie Sk\l{}odowska-Curie Individual Fellowships) with the Computer Vision and Graphics Group, Fraunhofer Heinrich-Hertz-Institut (HHI), Germany. His current research interests include 3D visual coding, processing, and communication. He is also serving as an Area Chair for IEEE ICME, an Associate Editor for \emph{IEEE Transactions on Image Processing}, \emph{IEEE Transactions on Consumer Electronics}, and \emph{IET Image Processing}. 
\end{IEEEbiography}
\begin{IEEEbiography}[{\includegraphics[width=1in,height=1.25in,clip,keepaspectratio]{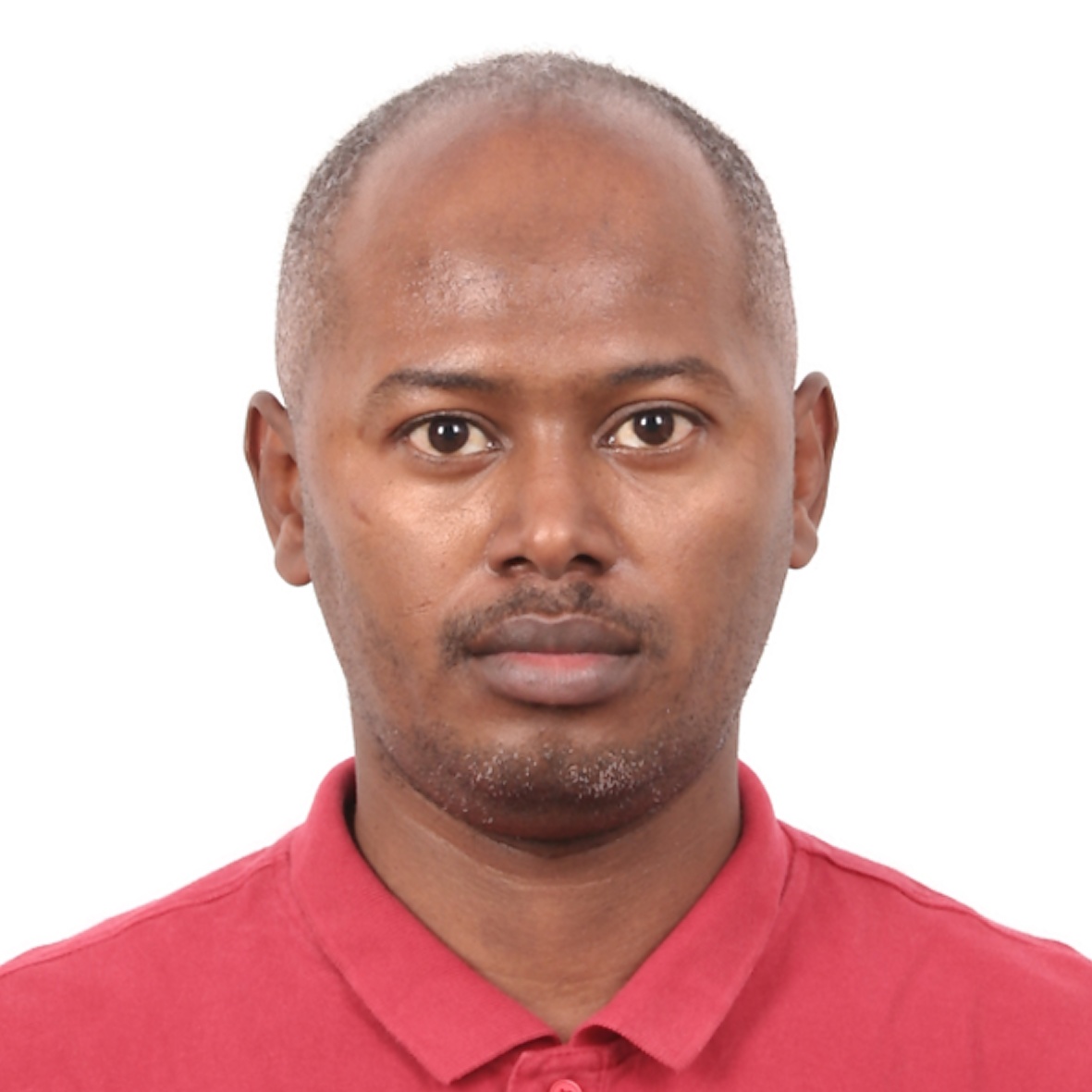}}]{Mohanad M. G. Hassan}
received the B.Sc. degree in Electronic Engineering in 2010 and the M.Sc. degree in Telecommunication Engineering in 2015, both from the University of Gezira, Sudan. He was a Teaching Assistant with the Department of EE and has been a Lecturer since 2016. He is currently pursuing the Ph.D. degree with the School of Control Science and Engineering, Shandong University, China. His research include semantic communication for 3D point clouds.
\end{IEEEbiography}
\begin{IEEEbiography}[{\includegraphics[width=1in,height=1.25in,clip,keepaspectratio]{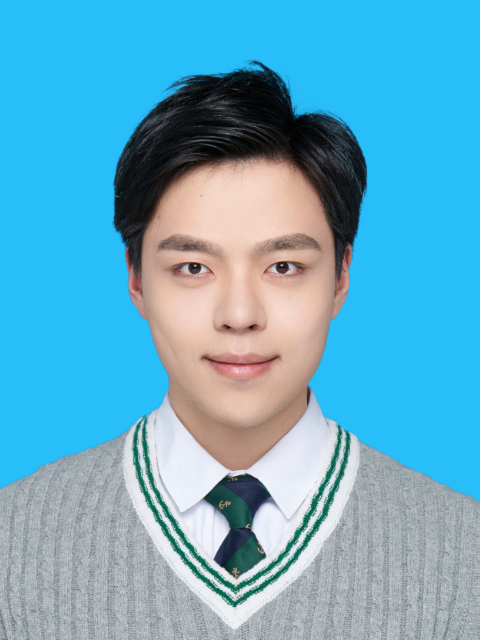}}]{Zejia Chen}received his BS degree in Communications Engineering from the School of Information and Electrical Engineering, Ludong University, China, in 2020, and his MS degree in Systems Science from the School of Automation, Qingdao University, China, in 2023. Since then, he has been pursuing a PhD degree at the School of Control Science and Engineering at Shandong University, China. His main research area is the semantic communication of 3D point clouds.
\end{IEEEbiography}
\begin{IEEEbiography}[{\includegraphics[width=1in,height=1.25in,clip,keepaspectratio]{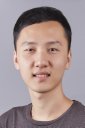}}]{Guanghui Zhang} is currently a full Professor at the School of Computer Science and Technology, Shandong University, China. He received his Ph.D. degree from the Department of Information Engineering of The Chinese University of Hong Kong in 2020, and an MS degree in Electronic Science and Technology from Peking University in 2016. From 2020 to 2022, he worked as a postdoctoral researcher at The Chinese University of Hong Kong, and then as a research assistant professor at The Hong Kong Baptist University. His research interest broadly lies in networking systems, multimedia systems, and machine learning.
\end{IEEEbiography}
\end{document}